%% file: main.tex
\documentclass[manuscript,nonacm,screen]{acmart}
\AtBeginDocument{%
  }
    
\usepackage{orcidlink}
\usepackage{algorithm}
\usepackage{algpseudocode}
\usepackage{multirow}
\usepackage{listings}
\usepackage{tcolorbox}
\usepackage{booktabs}
\usepackage{xcolor}
\usepackage{subcaption} 
\usepackage{graphicx}
\usepackage{caption}
\usepackage{pgfplots}
\pgfplotsset{compat=1.18} 
\usepackage{pgfplotstable} 
\usepackage{xcolor} 
\usepackage{ulem}

\input{macros}

\acmISBN{978-1-4503-XXXX-X/2018/06}

\begin{document}

\title{Can Code Language Models Learn Clarification-Seeking Behaviors? 
}

\author{Jie JW Wu\orcidlink{0000-0002-7895-2023}}
\affiliation{%
  \institution{Michigan Technological University, Houghton}
  \department{Department of Computer Science}
  \city{Houghton}
  \state{MI}
  \postcode{49931}
  \country{USA}
}
\email{jie.jw.wu@acm.org}

\author{Manav Chaudhary}
\affiliation{%
  \institution{Indian Institute of Information Technology (IIIT)}
  \country{India}}
\email{manav.chaudhary@research.iiit.ac.in}

\author{Davit Abrahamyan}
\affiliation{%
  \institution{University of California, San Diego (UCSD), La Jolla}
  \country{USA}}
\email{dabrahamyan@ucsd.edu}

\author{Arhaan Khaku}
\affiliation{%
  \institution{University of British Columbia, Kelowna}
  \city{Kelowna}
  \country{Canada}}
\email{arhaank@student.ubc.ca}

\author{Anjiang Wei}
\affiliation{%
  \institution{Stanford University, Stanford}
  \country{USA}}
\email{anjiang@stanford.edu}

\author{Fatemeh Fard\orcidlink{0000-0002-4505-6257}}
\affiliation{%
  \institution{University of British Columbia, Kelowna}
  \city{Kelowna}
  \country{Canada}}
\email{fatemeh.fard@ubc.ca}

\renewcommand{\shortauthors}{Wu et al.}

\newcommand{\jw}[1]{\textcolor{black}{{#1}}}
\newcommand{\FHF}[1]{\textcolor{black}{{#1}}}

\begin{abstract}

Large language models (LLMs) have demonstrated remarkable capabilities in code generation tasks. However, a significant gap remains between their output and problem-solving strategies of human developers. Unlike human developers, who allocate substantial time to disambiguating requirements through iterative dialogue, LLMs often generate code despite the ambiguities in natural language requriements, leading to unreliable solutions. Different from prior work, we study whether a Code LLM can be fine-tuned to learn clarification-seeking behavior. While recent work has focused on LLM-based agents for iterative code generation, we argue that the fundamental ability to recognize and query ambiguous requirements should be intrinsic to the models themselves, especially in the context of agentic AI where models and humans collaborate frequently. In this research, we present ClarifyCoder, a new framework with synthetic data generation and instruction-tuning that fine-tunes an LLM to identify ambiguities, if they exist, and request clarification before proceeding with code generation. Our approach consists of two main components: (1) a data synthesis technique that augments existing programming datasets with scenarios requiring clarification to generate clarification-aware training data, and (2) a fine-tuning strategy that teaches models to prioritize seeking clarification over immediate code generation when faced with incomplete or ambiguous requirements. We further provide an empirical analysis of integrating ClarifyCoder with standard fine-tuning for a joint optimization of both clarification-awareness and coding ability. Experimental results demonstrate that ClarifyCoder achieves a 63\% communication rate (a 40\% absolute increase) and a 52\% good question rate (30\% absolute increase) on ambiguous tasks, significantly improving the communication capabilities of Code LLMs through meaningful clarification dialogues, while maintaining code generation capabilities.

\end{abstract}

\setcopyright{none} 
\settopmatter{printacmref=false} 

\renewcommand\footnotetextcopyrightpermission[1]{}

\maketitle
\section{\jw{Introduction}}
Large language models (LLMs) \cite{vaswani2017, svyatkovskiy2020, wang2021, feng2020} have demonstrated outstanding capabilities in code generation from natural language requirements, as shown by systems such as Codex~\cite{chen2021evaluating}, AlphaCode~\cite{li2022competition}, and CodeGen~\cite{nijkamp2022codegen}. Given the natural language requirements, or prompt, as input, the LLMs perform code generation that outputs the precise code snippets~\cite{hassan2024rethinking,10.1145/3660810}. 

\anjiang{This previous sentence here is a bit odd: does not connect well with the later setences. You should think about making the flow natural, e.g., starting from current benchmarks focus on translating NL description to code, however, it can be ambiguous in real-world setting, which is different from PL. Or you could move it to next paragraph, after sentence ``ACQ remains underexplored in code
generation tasks''.} 
However, unlike programming languages that are designed to be precise and unambiguous, the coding tasks described in natural language are inherently ambiguous~\cite{hoare1969axiomatic}. For instance, suppose the coding task is ``\textit{Return list with elements incremented by a number.}''~\cite{wu2024benchmarking}. The description is ambiguous as it does not indicate the specific value of the number. For software engineers, they may ask questions such as ``\textit{could you clarify the specific value of the number in the requirement?}'' to disambiguate the description in this coding task. However, several studies ~\cite{wu2024benchmarking} have shown that LLMs directly generate code rather than asking clarification questions in these cases~\cite{miao2025clarigen,10.1145/3660810}. More specifically, in HumanEvalComm benchmark~\cite{wu2024benchmarking}--a benchmark for evaluating LLMs' communication ability, LLMs generate code outputs in over 63\% of ambiguous scenarios without seeking necessary clarifications. 

The disambiguation of coding tasks is not only necessary but also important to ensure that the generated code is reliable and trustworthy. In real-world problem-solving processes, people, including software engineers, rely on continuous dialog to resolve ambiguities, align mental models, and iteratively refine problem statements and their solutions~\cite{whitehead2007collaboration,pressman2005software,mistrik2010collaborative}. Based on the study in ~\cite{meyer2019today}, senior developers allocate only 10–20\% to coding, with the majority remainder devoted to code reviews (25\%), meetings (20\%), and system design (30\%). This suggests that developers spend a significant amount of time on task disambiguation, including requirements, design, solutions, and meetings to align the requirements with stakeholders. On the other hand, in the new era of Agentic Software Engineering, the role of intelligent agents is becoming more prominent, where the Software Agents and Human Developers are envisioned to work collaboratively~\cite{hassan2025agentic}.
In this context, we anticipate an increasing need and importance for intelligent agents to resolve ambiguities and ensure the trustworthiness of software engineering tasks~\cite{belcak2025small,hassan2025agentic} in a human-AI collaboration environment. 


To address the above issue, several works have been proposed to disambiguate requirements for code generation using an iterative LLM-based workflow. TICODER~\cite{lahiri2022interactive} proposes an interactive workflow for guided intent clarification through tests to generate more accurate code. TICODER interacts with the user in the form of formal specifications. ClarifyGPT~\cite{10.1145/3660810} is a multi-step LLM-based code generation workflow that performs intent clarifications using a combination of test generation, code check, reasoning, and refinement. Okanagan~\cite{wu2024benchmarking} is a multi-round structure with customized prompts for asking clarifying questions when needed in code generation tasks where tasks could be ambiguous. Recently, ClariGen~\cite{miao2025clarigen} has been proposed using three LLMs together with interaction with the user to answer clarifying questions for enhancing code generation.

In spite of continuous research efforts, existing work may still face challenges when applied to code generation, where requirements can be either ambiguous or well-specified. First, user interaction with clarifying questions is still enforced for clear coding tasks that could be completed without clarifications. ClarifyGPT, TICODER, and ClariGen explicitly require the user interaction module, where it is a \textit{must} that the method needs to interact with the user or LLM-based simulation. Different from the other three methods, Okanagan is set up to either generate code or ask clarifying questions if only clarification is needed (i.e., when the coding task is considered ambiguous). However, Okanagan tends to still ask questions that appear to be unnecessary for original problems that do not need questions (pass@1 dropped from 65\% down to 27\% for standard coding tasks). This indicates a need to strike a balance between asking unnecessary questions and truly needed questions. Second, all of the existing works use LLM-based workflows and tend to be more complicated and costly than a single LLM. Even if we don't count the LLM-simulation that provides answer to the clarifying questions, ClariGen and Okanagan needs 3 calls to LLMs with customized prompts for each call, while ClarifyGPT needs 4 calls together with other modules such as test input generation and code consistency checks.   

\begin{figure*}[h]
  \centering
  \includegraphics[width=\textwidth]{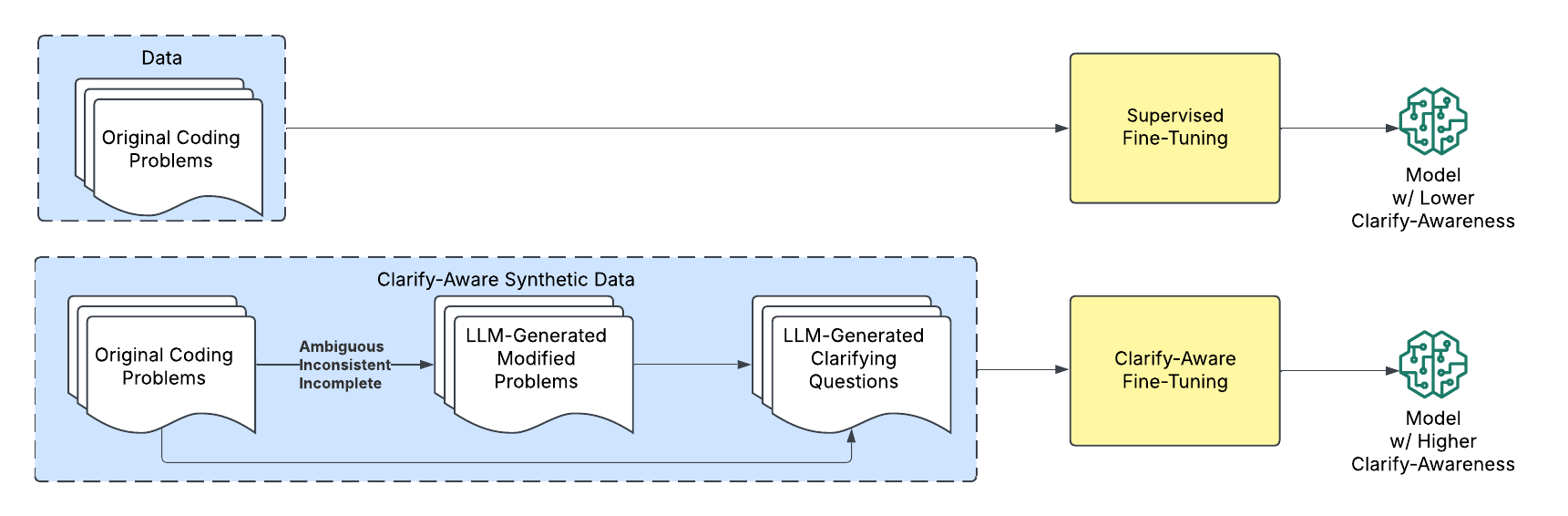}
  \caption{Overview of ClarifyCoder, including the synthetic data construction and clarification-aware fine-tuning.  
  }
  \label{fig:overall}
\end{figure*}

Our aim in this work is to explore the research question: \textit{Can Code LLMs learn clarification-seeking behavior?} In the real-world code generation task, the input coding task may be standard or ambiguous. If the input is standard, it may not be necessary to have user interaction each time. We want to understand whether it's possible to make Code LLMs learn the clarification-awareness behavior that human developers have by changing model weights using techniques such as fine-tuning. To the best of our knowledge, there is little work on empowering Code LLMs to get clarification-aware behaviors for code generation.  

In this research, we propose \textbf{ClarifyCoder}, a novel framework that includes synthetic data generation and instruction-tuning techniques to fine-tune a Code LLM to have clarification-seeking behavior. 
With the goal to enhance a LLM's clarification awareness in code generation, the framework includes 1) a new synthetic data generation approach for automatically
creating ambiguous problem descriptions and corresponding clarifying questions, and 2) a technique combining automated data synthesis with targeted instruction tuning for \textbf{clarification-aware} (or \textbf{clarify-aware}) code generation. As a result, ClarifyCoder can be integrated in the standard fine-tuning that satisfies both clarify-awareness and code generation capabilities.

We evaluate both the utility and clarify-awareness of our clarify-aware fine-tuning based on the proposed synthetic training data generation. Our experiments demonstrate that ClarifyCoder significantly improves LLMs' ability to recognize ambiguity and generate appropriate clarifying questions, including increases in communication rate from 24\% to 63\% and good question rate from 21\% to 52\%, with absolute increases of 40\% and 30\% respectively, \textit{while maintaining the code generation ability}. Besides, we also compared ClarifyCoder with training-free methods, including few-shot prompting and CoT reasoning, which showed significant lower clarify-awareness compared to ClarifyCoder. We further break down the results on different clarification categories and find that \textit{Incompleteness} is easiest for models to detect while \textit{Inconsistency} is most challenging. We further analyze the perplexity and entropy of the synthetic data generation in ClarifyCoder, and conduct a manual evaluation on LLM-based metrics to investigate the reliance of LLM-based metrics in evaluations. Finally, we investigate the impact of different synthetic data ratios and show that ClarifyCoder could be combined with standard fine-tuning to optimize both clarify-awareness and other utilities.  

Our main novelty in the existing literature is that, based on the experiments, the ClarifyCoder fine-tuning technique is able to empower an original LLM to ask questions (60\% of the time, more than doubled compared with the original LLM) to an ambiguous (or incomplete, or inconsistent) problem, and generate code (70\%+ pass rates) to a normal problem without sacrificing coding performance. Therefore, ClarifyCoder technique, to our knowledge, is one of the first works to achieve this capability in the literature. To advocate open science and full replication, we also released our data, full code on GitHub, and our model on HuggingFace. 

\textbf{Contributions. } To summarize, we make the following contributions:
\begin{itemize}
\item We propose a fine-tuning framework, ClarifyCoder, that combines synthetic data generation with targeted fine-tuning to enable clarify-aware code generation for a given LLM. The new synthetic data generation approach automatically generates ambiguous problem descriptions and corresponding clarifying questions. 
\item We conduct experiments to evaluate the effectiveness of ClarifyCoder in improving clarify-aware capabilities, while maintaining code generation capabilities of the models. This includes 1) comparison with different LLMs and training-free methods on clarify-awareness, 2) performance breakdown on clarification categories, 3) effectiveness of synthetic data generation, 4) manual evaluation of LLM-based metrics. In addition, we also evaluate the effect of different ClarifyCoder variants such as using different ratios of standard and synthetic data, and calculating loss on answer only vs both problem and answer.
\item A publicly available, open-source repository for full replication and future research on ClarifyCoder, at \href{https://github.com/jie-jw-wu/clarify-coder}{https://github.com/jie-jw-wu/clarify-coder}. The ClarifyCoder model weights are also released at \href{https://huggingface.co/jie-jw-wu/clarify-coder}{https://huggingface.co/jie-jw-wu/clarify-coder}.
\end{itemize}

\section{Background and Problem Formulation}
In this section, we introduce the background and formulation of the problem of clairfy-aware code generation.
\subsection{Backgrounds}

Language modeling forms the foundation of modern LLMs, particularly in decoder-only architectures like GPT-2 \cite{radford2019language} and GPT-3 \cite{brown2020language}. The core task involves autoregressive prediction of tokens in a sequence, where each token is predicted based on its preceding context. Formally, given a sequence of tokens \(x = \{x_1, \ldots, x_n\}\), the language model learns to predict each token \(x_i\) conditioned on its preceding tokens \(x_{<i}\). The training objective maximizes the log-likelihood:
\[
\mathcal{L}_{\text{LM}}(x) = \sum_{i=1}^n \log P(x_i \mid x_{<i}). \tag{6}
\]  
During generation, at each step $i$, the model samples token $x_i$ according to $\mathcal{L}_{\text{LM}}(x)$, which then becomes part of the context for sampling the next token $x_{i+1}$.

The development of LLMs typically involves two phases:
\begin{itemize}
    \item \textbf{Pretraining}: Models learn to predict next tokens across vast corpora, developing general language understanding and generation capabilities.
    \item \textbf{Instruction Tuning}: Models are fine-tuned to follow specific instructions and align with human preferences, transforming raw language capabilities into useful applications.
\end{itemize}

\begin{figure}[t]
\centering
\begin{lstlisting}[language=Python]

#Fixed scalar addition
def increment_list(last):
  return [x + 5 for x in lst]

#Parameterized increment
def increment_list(lst, num):
  return [x + num for x in lst]

#Position-dependent increment
def increment_list(last):
  return [x + i for i, x in enumerate(lst)]

#Type-conditional increment
def increment_list(last):
  return [x+1 if isinstance(x, int) else x \
             for x in lst]
\end{lstlisting}
\caption{Ambiguous implementations for the problem statement ``Return list with elements incremented by a number''. Variants demonstrate alternative interpretations of the problem.\anjiang{I added the indents and split the last function into more lines}}
\label{fig:ambig_code}
\end{figure}
Our research focuses on extending instruction tuning for code generation tasks to incorporate clarification-aware behavior. Current instruction-tuned models often generate code even when faced with ambiguous or incomplete specifications, a problem documented in \cite{wu2024benchmarking}. We propose a dual objective: 
    \begin{itemize}
        \item Maintain the model's code generation capabilities for well-specified tasks.
        \item Enable the model to recognize ambiguity and request clarification when necessary. 
\end{itemize}
    
We specifically target instruction tuning for this capability due to its efficiency and widespread adoption in language model development. This approach allows us to modify model behavior without the computational and data requirements of full pretraining, while still achieving significant improvements in clarify awareness.

\begin{table}[t]
\centering
\caption{Comparison of responses of DeepSeek Coder and ClarifyCoder 
for an ambiguous problem statement. ClarifyCoder is fine-tuned based on 
DeepSeek Coder and demonstrates the ability to ask clarifying questions 
to disambiguate coding problems. This example is from an \textit{incomplete} 
problem in HumanEvalComm, modified from HumanEval/43 in the HumanEval benchmark.}
\label{tab:example}

\scriptsize 
\begin{tabular}{p{4cm}p{4cm}p{4cm}} 
\toprule
\textbf{Ambiguous Problem Statement} & 
\textbf{DeepSeek Coder Response} & 
\textbf{ClarifyCoder Response} \\
\midrule

\begin{lstlisting}[language=Python,basicstyle=\ttfamily\scriptsize]
def candidate(l):
    """
    candidate takes a list of 
    integers as input.
    it returns True if there 
    are two distinct elements 
    in the list that
    """
\end{lstlisting}
&
\begin{lstlisting}[language=Python,basicstyle=\ttfamily\scriptsize]
def candidate(l):
    seen = set()
    for num in l:
        if num in seen:
            return True
        seen.add(num)
    return False
\end{lstlisting}
&
It seems like your function description is incomplete. 
Could you please provide more details about what the 
function should return if there are two distinct 
elements in the list? \\

\bottomrule
\end{tabular}
\end{table}

\subsection{Problem Formalization}

\label{sec:formalism}

Following the annotation of~\cite{kobalczyk2025active}, let $\Sigma$ denote the space of natural language. Within the context of AI-based software engineering assistance,
we define a problem statement of code generation $S \in \Sigma$ as a natural language instruction for a model to generate a code solution $h \in \Sigma$. The unknown set of ground-truth code solutions $H^* \subset \Sigma$ refers to the code solutions that pass the test cases or meet the users' intents. We assume that the problem statement, $S$, can be decomposed into two parts: a set of requirements $R$ that any $h \in H^*$ should satisfy, and any additional contextual information $C$ such as the system instruction in prompt that could affect the preference of different code solutions.

\FHF{We adopt the taxonomy of clarification types (ambiguous, inconsistent, and imcomplete problems) from previous works~\cite{wu2024benchmarking}. }
\jw{The taxonomy was designed by following clarification types based on both the literature in Requirement Engineering (RE) and understanding of how feasible can the RE concepts can be applied to problems in HumanEval:}

\begin{itemize}
    \item {\textit{Ambiguous Problem}}: \jw{Some statements in the problem descriptions could be ambiguous and correspond to different concepts.}
    \item {\textit{Inconsistent Problem}}: Some statements in the problem descriptions show conflict or inconsistency between each other.
    \item {\textit{Incomplete Problem}}: Some concepts or conditions are missing in the problem descriptions.
\end{itemize}

For a coding problem $S \in \Sigma$ that is \textit{ambiguous}, or \textit{inconsistent}, or \textit{incomplete}~\cite{dermeval2016applications,tukur2021requirement,wu2024benchmarking},  \textbf{the goal} of this research is to have the model ask clarifying questions so that the model, when given the answers to the clarifying questions in interaction, generates correct code $h 
\in H^*$. One assumption we make in this work is that the clarifying questions are answered correctly.  This enables a systematic analysis of disambiguation or under-specification in programming tasks. 

\jw{Inspired by~\cite{kobalczyk2025active}, we further develop this taxonomy of clarification types into formal definitions for code generation tasks.
Let $H^*$ denote the ground-truth code solutions that are correct. Let $H$ denote the solution space that comprises all syntactically valid implementations that meet the requirements $R$ of the problem statement.} \anjiang{that could match the natural language description? I don't think the definition of H is clear. There are infinitely many syntactically correct programs}:
\begin{equation}
    H := \{h : h \vdash R\}
\end{equation}
We provide definitions of ambiguous, incomplete and inconsistent problems as follows.
\begin{definition}[Ambiguous Problem~\cite{kobalczyk2025active}]
A coding problem $S$ exhibits ambiguity if $H^* \subset H$ ($H^*$ is a proper subset of $H$), i.e., \begin{equation}
\exists h \vdash R\quad \text{s.t.} \quad h \not\in H^*
\end{equation}
\end{definition}

Consider the canonical example:

\begin{lstlisting}[language=Python]
def incr_list(l: list):
"""Return list with elements incremented by a number."""
\end{lstlisting}

\jw{This specification permits multiple valid interpretations $H = \{h_{\text{1}}, h_{\text{2}}, h_{\text{3}}, h_{\text{4}},...\}$. However, the ground truth code solution $H^*$ for this problem statement is only a proper subset of $H$.} In this example, the ambiguity stems from an underspecified number to increment by. See also another example in Figure~\ref{fig:ambig_code}. Given this ambiguous problem, the goal of this model is to ask clarifying questions to disambiguate the coding problem. When the clarifying question(s) are answered, with this additional information, the model attempts to generate the correct code $h 
\in H^*$.

\begin{definition}[ Incomplete Problem]
A coding problem $S$ is incomplete if:
\begin{equation}
    \exists r_{\text{missing}} \not\in R \quad \text{s.t.} \quad H^{*} = \{h : h \vdash R \cup \{r_{\text{missing}}\}\} \subset H
\end{equation}
\end{definition}
A collary of this definition is that an \textit{incomplete} problem must also be \textit{ambiguous}.
For coding problem statements, examples of missing requirements might include:
\begin{itemize}
    \item Time complexity constraints: $r_{\text{time}} := O(\log n)$
    \item Error handling: $r_{\text{error}} := \text{Throw ValueError for null inputs}$
\end{itemize}
Table~\ref{tab:example} shows another real example of \textit{incomplete} problem and different models' responses.

\begin{definition}[ Inconsistent Problem]
A coding problem with specifications contains conflicting requirements and is thus inconsistent if $ H = \{h : h \vdash R\}=\emptyset$, i.e.,
\begin{equation}
    \exists r_i,r_j \in R \quad \text{s.t.} \quad \nexists h \in \Sigma \quad h \vdash r_i \land h \vdash r_j
\end{equation}
\end{definition}

An example of inconsistent problem include the following two requirements that contradict each other:
\begin{align*}
    r_1 & := \text{Return list with elements incremented by 1. (in docstring) } \\
    r_2 & := \text{> > > incr\_list ([1, 2, 3]) [3, 4, 5] (indicating incremented by 2)} 
\end{align*}

\section{Clarify-Aware Instruction Tuning}
\label{sec:clarify_tuning}

This section details our methodology for enhancing Large Language Models with clarify-aware instruction tuning, specifically tailored for code generation tasks. As illustrated in Figure~\ref{fig:overall}, given a constructed dataset comprising both original and clarification-focused examples, we explore several fine-tuning strategies. The goal is to enable the LLM with the capacity to not only generate code adhering to functional specifications but also to proactively elicit necessary clarifications to resolve ambiguities and address potential incompleteness in problem descriptions.

\subsection{Baseline Instruction Tuning}
We first establish a baseline by employing standard instruction tuning on the original programming dataset, denoted as $D^{og}$\anjiang{What does ``original programming dataset'' mean? Do you mean that these problems are clearly specified without the need to ask questions?} \anjiang{Also, one big question I have is that, if the users provide test cases, and have a feedback mechanism, then it seems that ambiguous specification is also fine, since the test case feedback already tells what's going on. Do you consider adding experiments to demonstrate evaluation on pass@1 without test case feedback?}. This dataset consists of pairs $(p, a)$, where $p$ represents a programming task description and $a$ represents a corresponding code solution that satisfies all specified test cases.  Following conventional practice, we fine-tune the LLM by minimizing the cross-entropy loss, focusing on generating the code solution $a$ given the task description $p$:

\begin{equation}
    \mathcal{L}^{og}= -\sum_{t=1}^{|a|} \log P(a_t|a_{<t},p)
    \label{eq:original_loss}
\end{equation}

While this standard tuning approach enhances the LLM's ability to generate correct code, it inherently neglects the clarify-aware dimension. By training solely on complete, unambiguous examples, the model develops a systematic disadvantage in its capacity to recognize and address underspecification in problem statements through targeted clarification requests.

\subsection{Clarify-Aware Instruction Tuning}
To address the limitations of standard instruction tuning, we introduce a novel approach leveraging a dedicated clarify-aware dataset, $D^{clarify}$. Each sample $(p^{clarify}, a^{clarify})$ in $D^{clarify}$ consists of a specially crafted instruction $p^{clarify}$ and a corresponding clarification output $a^{clarify}$. The instruction $p^{clarify}$ is designed to elicit a clarifying response from the LLM and incorporates:

\begin{enumerate}
    \itemsep0em
    \item  A system prompt: Guiding the LLM to either ``ask a clarifying question'' or ``generate code'' based on the task's clarity.
    \item  A programming task description: Potentially containing ambiguities or underspecifications.
\end{enumerate}

The clarification output $a^{clarify}$ provides the appropriate clarifying question(s) designed to resolve any perceived ambiguities or incompleteness in $p^{clarify}$.

To maintain simplicity and facilitate direct comparison, we fine-tune the LLM on $D^{clarify}$ using the same cross-entropy loss minimization framework as in standard instruction tuning:

\begin{equation}
    \mathcal{L}^{clarify}= -\sum_{t=1}^{|a^{clarify}|} \log P(a^{clarify}_t|a^{clarify}_{<t},p^{clarify})
    \label{eq:clarify_loss}
\end{equation}

By minimizing $\mathcal{L}^{clarify}$, we aim to equip the LLM with the ability to discern when clarification is necessary and to generate effective questions that elicit the missing or ambiguous information required for accurate code generation.

\subsection{Combined Instruction Tuning}
To synergistically leverage the strengths of both standard and clarify-aware instruction tuning, we propose a combined training approach. We create a unified dataset $D^{all}$ by merging $D^{clarify}$ and $D^{og}$:

\begin{equation}
    D^{all}= D^{clarify} \cup D^{og}
\end{equation}

We then perform instruction tuning on $D^{all}$ by minimizing the combined loss function $\mathcal{L}^{all}$:

\begin{equation}
    \mathcal{L}^{all}= -\sum_{t=1}^{|a^{all}|} \log P(a^{all}_t|a^{all}_{<t},p^{all})
    \label{eq:combined_loss}
\end{equation}

This approach allows the LLM to simultaneously learn to generate correct code from complete specifications and to proactively seek clarification when faced with underspecified tasks. The combined loss function encourages a balanced learning process, fostering both code generation proficiency and clarify-aware reasoning.

\subsubsection{Dataset Balancing and Sampling Strategies}
The relative sizes of $D^{clarify}$ and $D^{og}$ can potentially influence the performance of the combined training approach. We introduce the ratio $r = |D_{\text{clarify}}| / (|D_{\text{og}}|+|D_{\text{clarify}}|)$ to control the balance between the two datasets. We systematically investigate the impact of varying $r$ on the resulting model's ability to generate code and ask clarifying questions.

Furthermore, we explore different sampling strategies for combining the two datasets:

\begin{itemize}
    \itemsep0em
    \item \textbf{Uniform sampling:}  Samples are drawn uniformly from the combined dataset $D^{all}$.
    \item \textbf{Oversampling:}  The smaller dataset ($D^{clarify}$ or $D^{og}$, depending on $r$) is oversampled by randomly duplicating samples until the desired ratio $r$ is achieved. This prioritizes the learning of the less represented task (either clarification or code generation).
    \item \textbf{Downsampling:}  The larger dataset is downsampled by randomly removing samples until the ratio $r$ is satisfied.
\end{itemize}

In the evaluation, we used downsampling because we did not find much difference compared with oversampling. The detailed steps of the combined instruction tuning process are summarized in Algorithm~\ref{alg:instruction_tuning}.

\begin{algorithm}[H]
    \caption{Combined Instruction Tuning for Code Generation}
    \label{alg:instruction_tuning}
    \begin{algorithmic}[1]
        \Require Pre-trained LLM
        \Require $D_{\text{og}}$: Standard instruction tuning dataset
        \Require $D_{\text{clarify}}$: Clarify-aware instruction tuning dataset
        \Require $r$: Ratio $|D_{\text{clarify}}| / (|D_{\text{og}}|+|D_{\text{clarify}}|)$
        \Ensure Instruction-tuned LLM

        \State $D^{\text{all}} \gets $ Combine $D^{\text{og}}$ and $D^{\text{clarify}}$ according to ratio $r$ and chosen sampling strategy
        \For{each sample $s \in D^{\text{all}}$}
            \State Optimize the LLM on $s$ by minimizing $\mathcal{L}^{\text{all}}$ (Equation \ref{eq:combined_loss})
        \EndFor
        \State \Return Instruction-tuned LLM
    \end{algorithmic}
\end{algorithm}

\section{Clarify-Aware Data Construction}

This section details the methodology used to generate modified coding problem descriptions and their corresponding clarifying questions. Our objective was to produce problem statements characterized by ambiguity, inconsistency, or incompleteness, and their corresponding clarifying questions, thereby creating a robust dataset for training ClarifyCoder.

\subsection{Dataset}

We utilize the APPS dataset, which comprises of 10,000 coding problems sourced from diverse open-access platforms like Codeforces and Kattis. Each problem is accompanied by multiple test cases and human-written solutions, covering a range of complexities from introductory to collegiate competition levels, with an average description length of 293.2 words.

Given the constraints of our budget, we opted to employ Google Gemini due to its strong performance and accessibility via a free research API. This API facilitates smooth integration into our scripts, enabling efficient model calls and robust error management.

\subsection{Generating Modified Problems}

We instruction-tuned Gemini to modify original problem statements, introducing ambiguity, incompleteness, or inconsistency. Gemini was prompted using a chain-of-thought approach with the instructions shown in Table~\ref{tab:prompts}, where each instruction targets a specific type of modification.

\begin{table}[t]
    \centering
    \caption{Gemini Prompts for Problem Modification and Question Generation}
    \label{tab:prompts}
    \small
    \begin{tabular}{p{1.5cm}p{7cm}}
        \hline
        \textbf{Problem Type} & \textbf{Prompts} \\
        \hline
        Ambiguous & \textbf{Modification:} Rewrite to introduce ambiguity by creating multiple valid interpretations or leaving key details unspecified.\\
        & \textbf{Question Gen:} Identify points of ambiguity and formulate clarifying questions that resolve them. \\
        \hline
        Incomplete & \textbf{Modification:} Rewrite to create incompleteness by omitting key concepts or conditions essential for solving. \\
        & \textbf{Question Gen:} Identify points of incompleteness and formulate clarifying questions to resolve them. \\
        \hline
        Inconsistent & \textbf{Modification:} Rewrite to introduce inconsistency by incorporating conflicting statements. \\
        & \textbf{Question Gen:} Identify points of inconsistency and formulate clarifying questions to provide a consistent interpretation. \\
        \hline
    \end{tabular}
\end{table}

We instruction-tune the Gemini model to modify original problem statements through a chain-of-thought knowledge-infused prompting strategy. This includes modifications specific to each modification category. Table~\ref{tab:prompts} illustrates the instructions we used to make the model generate modified problems from their original counterparts.

\subsection{Generating Clarifying Questions}
Following problem modification, we generated clarifying questions by prompting Gemini with both the modified and original problem descriptions. We use the prompts outlined in Table~\ref{tab:prompts}, tailoring the instruction to the specific type of problem modification.

We ensure that the generated clarifying questions must focus solely on the modified problem descriptions, without referencing the original versions, since the original version would never be available to the finetuned model. We provide the original questions in the instruction so that the generated clarifying questions are relevant to the problem at hand, i.e., generating code, ensuring that the model does not ask irrelevant questions. We use the following instruction at the end of our prompt to achieve this:

\begin{itemize}
    \item \textit{Ensure that each question targets a specific point to achieve clarity similar to the original problem. When generating these questions, do not reference or mention the original problem description in any way. Frame the clarifying questions as if you have only seen the modified problem description, without acknowledging the existence of the original version. } \anjiang{Should the detailed prompt go into the appendix?}
\end{itemize}

\subsection{Dataset for Clarify-Aware Fine-Tuning}
As aforementioned, we use the APPS dataset for standard instruction tuning, which comprises of 10,000 coding problems sourced from diverse open-access platforms like Codeforces and Kattis. Each problem is accompanied by multiple test cases and human-written solutions, covering a range of complexities from introductory to collegiate competition levels, with an average description length of 293.2 words.

As for the dataset of clarify-aware instruction tuning, we performed the data construction on top of APPS dataset. As a result, we generated 29,896 samples as the clarify-aware dataset. The number of samples for each of the three categories (ambiguous, inconsistent, incomplete) is roughly 10,000.

In the final step, we consolidate the modified problems and their corresponding clarifying questions into a unified dataset tailored for fine-tuning the language model. This dataset contains three primary fields: \texttt{problem}, \texttt{answer}, and \texttt{clarification category}.

\section{Experimental Design}
This section describes the research questions, dataset, models and measurements of this work. 
\subsection{Research Questions}
 To evaluate ClarifyCoder, we address the following research questions:

\anjiang{RQs too long. The 3 RQs should be very clear and simple. You can put text for explanation later.}

\begin{itemize}
\item \textbf{RQ1:} \textit{To what extent can clarify-aware fine-tuning improve a model's ability to ask effective clarifying questions?} This question examines the impact of our clarify-aware fine-tuning techniques on the model’s ability to generate meaningful and contextually relevant clarifying questions. We assess whether models trained with our method generate clarifications that are both necessary and helpful using different evaluation metrics.
\item \textbf{RQ2:} \textit{How does ClarifyCoder perform in problems of different clarification categories, compared with other models? } This question investigates the effectiveness of ClarifyCoder across various clarification categories. We analyze its performance relative to other models to determine whether it excels in generating relevant clarifications for different types of ambiguity.  
\item \textbf{RQ3:} \textit{How effective are our proposed data synthesis methods in introducing ambiguity into problems while maintaining structured answers?} This question evaluates whether our synthetic data generation approach produces the intended pattern of ambiguous problems paired with consistent clarifying questions, as measured by perplexity and entropy metrics.

\item \textbf{RQ4:} \textit{How does clarify-aware fine-tuning compare to training-free methods in improving clarification awareness?} We perform experiments using in-context learning and chain-of-thought prompting and compare the results achieved through them to those achieved with the proposed Clarify-Aware fine-tuning approach.
\end{itemize}

\subsection{Dataset (HumanEval and HumanEvalComm)}
We tested different models on coding tasks with both clear requirements and ambiguous requirements. For clear requirements, we used the widely used HumanEval benchmark. For ambiguous requirements, we opted to use HumanEvalComm~\cite{wu2024benchmarking}, a benchmark dataset for evaluating communication skills in LLMs during code generation. HumanEvalComm is built by modifying the widely used HumanEval dataset to include incomplete, inconsistent, and ambiguous requirements. One of the authors in this paper, did a thorough exam on the dataset to make  corrections if needed. We found that in a small number of cases, the problem description was not modified enough to result in an LLM asking a question, leading to a lower Communication Rate than expected. For those questions, we modified the wording to make sure that they would result in possible clarifying questions. Modifications included rewording certain problems and adding ambiguous, incomplete, or inconsistent variations. The updated dataset maintains a similar distribution of question types with slight adjustments in some categories. This refined version aims to provide a more robust evaluation of models' ability to ask clarifying questions. The updated dataset is checked into the HumanEvalComm dataset on GitHub~\cite{wu2024benchmarking} and HuggingFace~\cite{humanevalcomm} after careful verifications.

\subsection{Models}
We evaluate five widely used LLMs: three open-source instruction-tuned Code LLMs, one open-source instruction-tuned general LLM, and our ClarifyCoder.

\textbf{CodeLlama} (Instruction-tuned, 13B)~\cite{roziere2023code} is an open-source LLM by Meta built on Llama 2, widely used for coding tasks with strong HumanEval performance. \textbf{DeepSeek Coder} (Instruction-tuned, 7B)~\cite{guo2024deepseek} is trained on 87\% code and 13\% natural language across 2 trillion tokens. It ranks in the top 5 on the Big Code Models Leaderboard~\cite{bigcodemodelshf}. \textbf{DeepSeek Chat} (Instruction-tuned, 7B)~\cite{bi2024deepseek} is a general-purpose LLM trained on 2 trillion tokens. We include it to compare with DeepSeek Coder and evaluate whether more natural language training improves communication skills. \textbf{CodeQwen1.5 Chat} (Instruction-tuned, 7B)~\cite{bai2023qwen} is trained on 3 trillion tokens of code data and employs group query attention (GQA) for efficient inference. It also ranks in the top 5 on the Big Code Models Leaderboard. \textbf{ClarifyCoder} (Our model, based on DeepSeek Coder 7B) is fine-tuned with our proposed technique to improve clarification capabilities.

We limit our evaluation to instruction-tuned models since foundation models without instruction tuning are not suitable for our evaluation task, which requires understanding instructions to either generate code or ask clarifying questions.


\subsection{Evaluating Utility and Clarify-Awareness}
To comprehensively evaluate the performance of our proposed clarify-aware instruction tuning approach, we adopt a suite of metrics inspired by the HumanEvalComm benchmark \citep{wu2024benchmarking}:

\begin{itemize}
    \itemsep0em
    \item \textbf{Pass@1:} The percentage of generated code solutions that pass all test cases on the first attempt. Pass@1 in HumanEval refers to code generation ability, while Pass@1 in HumanEvalComm means the Pass@1 in \textit{post-clarification }(code generation in 2nd round)  \textit{after} the LLM has engaged in clarification dialog and received answers to its clarifying questions, measuring the code ability in clarification process.
    \item \textbf{Test Pass Rate:} Test Pass Rate is defined as the proportion of successfully passed test cases in relation to the total number of test cases. Like Pass@1, Test Pass Rate in HumanEvalComm also refers to the pass rate in \textit{post-clarification}. 
    \item \textbf{Communication Rate (Comm. Rate):} The frequency with which the LLM chooses to ask clarifying questions rather than directly generating code. This indicates the model's awareness of task ambiguity. We used a LLM-based metric, prompting ChatGPT 4o-mini to output 0 (no question) or 1 (asked question) given a model's response.  
    \item \textbf{Good Question Rate (Good Q. Rate):} The percentage of clarifying questions that are relevant and helpful in resolving the ambiguity or incompleteness of the task description.  This metric assesses the quality of the generated questions. If the Comm. Rate is 1 (model response is question), we used a LLM-based metric, prompting ChatGPT 4o-min, to output 0 (not good question) or 1 (good question).
\end{itemize}

By considering these metrics in conjunction, we aim to provide a holistic assessment of the LLM's ability to generate correct code, proactively address ambiguities, and formulate meaningful clarification requests. The effectiveness of our approach will be judged by its ability to simultaneously improve test pass rates and generate high-quality clarifying questions.

\section{Results}

This section introduces the experimental results, analysis, and findings for each RQ.

\subsection{Clarification Competency of Code LLMs}

\begin{table}[t]
\centering
\renewcommand{\arraystretch}{1.2} 
\setlength{\tabcolsep}{8pt} 
\small 
\begin{tabular}{p{2.5cm}cccc}
\toprule
\multirow{2}{*}{\textbf{Model}} & \multicolumn{2}{c}{\textbf{HumanEvalComm}} & \multicolumn{2}{c}{\textbf{HumanEval}} \\
\cmidrule(lr){2-3} \cmidrule(lr){4-5}
& \textbf{Comm.} & \textbf{Good Q.} & \textbf{Pass@1} & \textbf{Test Pass} \\
& \textbf{Rate} & \textbf{Rate} & & \textbf{Rate} \\
\midrule
CodeLlama & 4.38\% & 3.93\% & 29.27\% & 41.63\% \\
CodeQwen1.5 Chat & 0\% & 0\% & \underline{\textbf{76.22\%}} & \underline{\textbf{84.95\%}} \\
DeepSeek Chat & \textbf{28.25\%} & \textbf{23.03\%} & 34.76\% & 53.00\% \\
DeepSeek Coder & \textbf{24.12\%} & \textbf{21.29\%} & \textbf{73.62\%} & \textbf{82.36\%} \\
ClarifyCoder {\tiny (w/ DS Coder)} & {\underline{\textbf{57.42\%}}} & {\underline{\textbf{47.68\%}}} & \textbf{73.01\%} & \textbf{82.00\%} \\
ClarifyCoder {\tiny (w/ DS Coder, loss=answer only)} & \underline{\textbf{63.61\%}} & \underline{\textbf{51.93\%}} & \underline{\textbf{74.85\%}} & \underline{\textbf{83.29\%}} \\
\bottomrule
\end{tabular}
\caption{\jw{Results on \textit{HumanEvalComm} and \textit{HumanEval} benchmarks for different LLMs. \underline{\textbf{First-place}} and \underline{\textbf{Second-place}} results are underlined and bolded, while \textbf{top 4} results are bolded. } \jw{"loss=answer only" denotes the loss during fine-tuning, calculated only on the answers rather than on both the problems and the answers.}
} 
\label{tab:experiment_performance_results}
\end{table}

\begin{figure}
  \centering
  \begin{subfigure}[b]{0.48\textwidth}
    \centering
    \includegraphics[width=\linewidth,height=0.3\textheight]{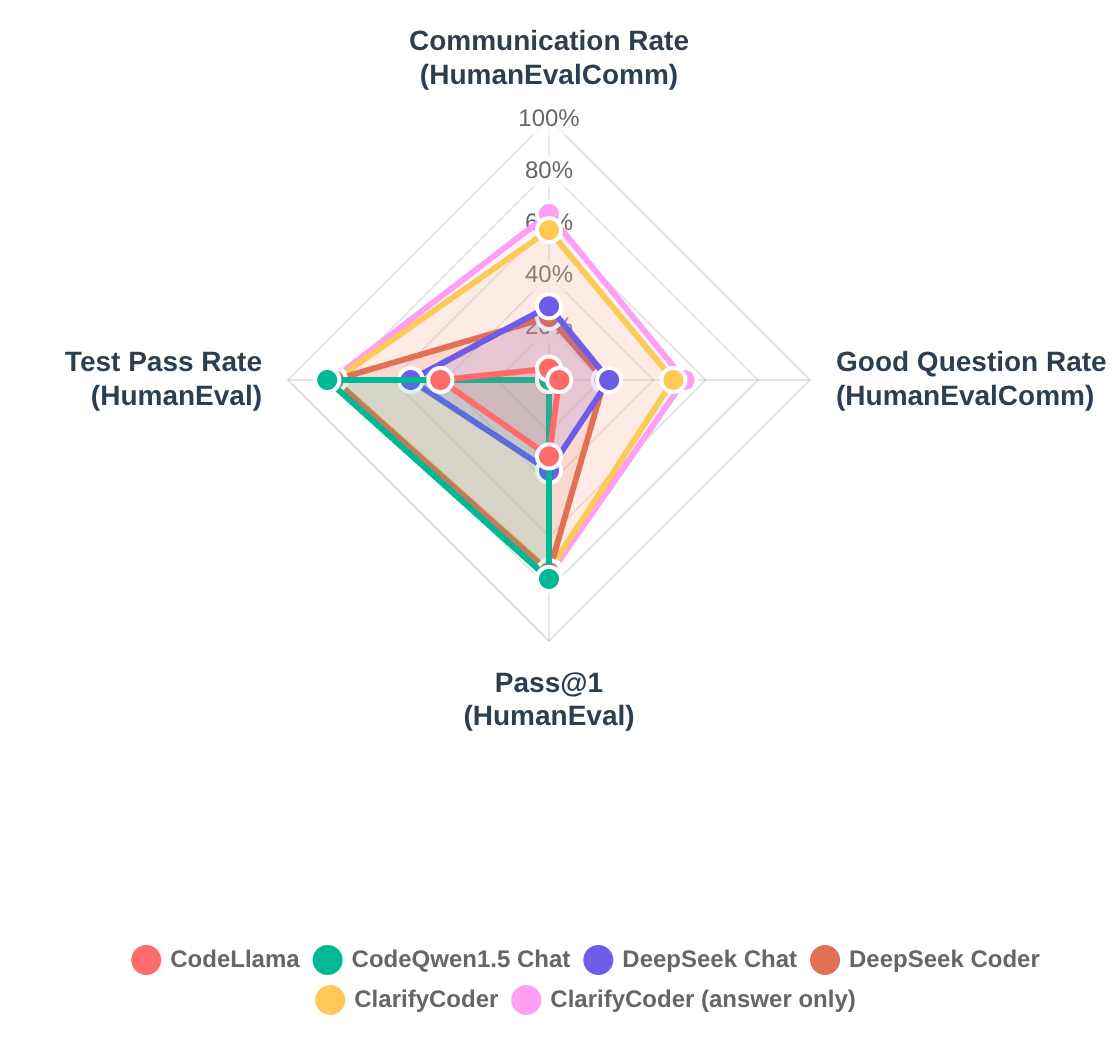}
    \caption{Multi-dimensional performance profile for clear (HumanEval) and ambiguous requirements (HumanEvalComm).}
    \label{fig:left}
  \end{subfigure}%
  \hfill
  \begin{subfigure}[b]{0.48\textwidth}
    \centering
    \includegraphics[width=\linewidth]{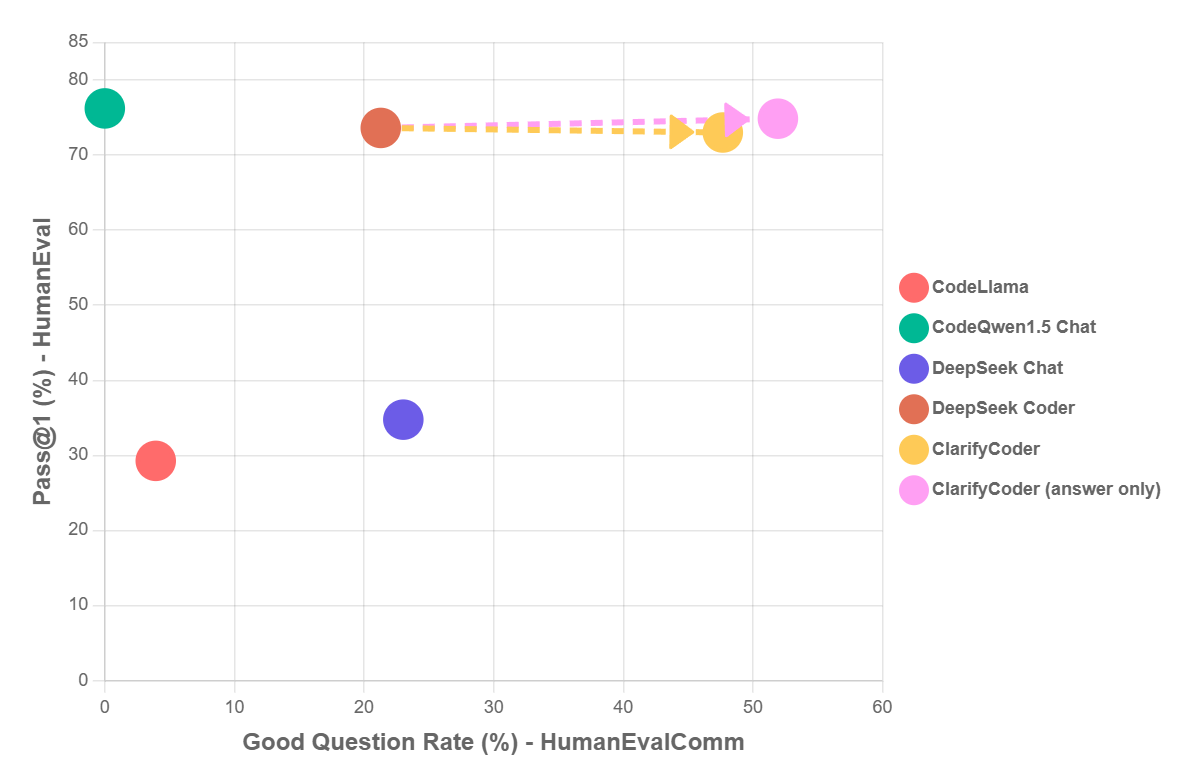}
    \caption{a scatter plot of the trade-off between clarification-seeking ability and code generation performance. Dashed arrow represents the impact of Clarification-aware fine-tuning on the original DeepSeek Coder model.}
    \label{fig:right}
  \end{subfigure}
  \caption{\jw{Comprehensive performance analysis of ClarifyCoder and baseline LLMs.} \jw{"loss=answer only" denotes the loss during fine-tuning, calculated only on the answers rather than on both the problems and the answers.}}
  \label{fig:two-subfigs}
\end{figure}

\begin{table}[t]
\caption{\jw{Results of different methods on HumanEval benchmark (well-specified coding tasks). 
Pass@1 and Test PR measure code generated in the first (only) round.  
The top two results are bolded. Pass rates are considered unavailable for methods where user interaction to perform clarification is enforced.  }}
\centering
\resizebox{0.85\textwidth}{!}{
\begin{tabular}{l l c c c c}
\toprule
\textbf{Method} & \textbf{Method Type} & \textbf{\# LLM calls} & \textbf{Interaction Enforced} & \textbf{Pass@1} & \textbf{Test Pass Rate} \\
\midrule
TICODER & LLM Workflow & 2 & Yes & N/A & N/A \\ 
ClarifyGPT & LLM Workflow & 4 & Yes & N/A & N/A \\  
ClariGen & LLM Workflow & 3 & Yes & N/A & N/A \\ 
Okanagan{\tiny (Base=ChatGPT)} & LLM Workflow & 3 & No & 27.45\% & 33.45\%  \\ 
Okanagan{\tiny (Base=DS Coder)} & LLM Workflow & 3 & No & 21.25\% & 24.3\%  \\ 
DeepSeek Chat & Single LLM & 1 & No & 34.76\% & 53.00\% \\ 
DeepSeek Coder & Single LLM & 1 & No & \textbf{73.62\%} & \textbf{82.36\%} \\ 
ClarifyCoder {\tiny (w/ DS Coder)} & Single LLM & 1 & No & 73.01\% & 82.00\%  \\ 
ClarifyCoder {\tiny (w/ DS Coder, loss=answer only)} & Single LLM & 1 & No & \textbf{74.85\%} & \textbf{83.29\%}  \\ 
\bottomrule
\end{tabular}
}
\label{tab:human_eval_results_compare}
\end{table}

To address RQ1, we evaluate the effectiveness of clarify-aware fine-tuning in improving a model's ability to generate meaningful clarifying questions. We compare our proposed ClarifyCoder with existing code generation models on the HumanEvalComm benchmark, which specifically tests performance on problems requiring clarification. 

\jw{Table \ref{tab:experiment_performance_results} displays the numerical results on \textit{HumanEvalComm} and \textit{HumanEval} benchmarks for different LLMs and ClarifyCoder. We included two variants of ClarifyCoder: one where loss during fine-tuning is calculated on both the problem and answer, and another one, with donation "loss=answer only", where loss is calculated solely on the answer.} From Table \ref{tab:experiment_performance_results}, we observe significant improvements in communication rate with both variants of ClarifyCoder. ClarifyCoder achieves a substantially higher communication rate of 63.61\% (for with answer only), more than doubling the rate of the original DeepSeek Coder (24.12\%) and the second-best model DeepSeek Chat (28.25\%). This demonstrates that our technique effectively improves the model's ability to recognize when clarification is needed. ClarifyCoder similarly excels in generating high-quality clarifying questions, with 51.93\% of its questions being evaluated as good, compared to 23.03\% for DeepSeek Chat. This indicates that our approach not only increases the frequency of clarifications but also their question quality.




\jw{Figure \ref{fig:two-subfigs}(a) shows a radar chart that shows each model's capability on four metrics regarding coding performance and clarification-seeking ability.  Figure \ref{fig:two-subfigs}(b) shows a scatter plot of the fundamental trade-off relationship between asking quality clarifying questions and generating correct code, with most models clustered in suboptimal regions. Comparing the multi-dimensional performance profile and the performance trade-off visualization in Figure \ref{fig:two-subfigs}, we observe that the two variants of ClarifyCoder experience very small performance degradation for clear requirements on HumanEval benchmark, while showing significant improvements in communication rates and Good Question Rates for ambiguous requirements on HumanEvalComm benchmark. This represents a good trade-off and advantage for adopting our clarification-aware fine-tuning. This suggests that ClarifyCoder effectively enables the models (in our case, DeepSeek Coder) to recognize when clarification is needed \textit{without} sacrificing code generation capabilities. By enhancing the model's ability to recognize ambiguity and generate relevant clarifying questions, ClarifyCoder demonstrates a more interactive and human-like approach to code generation tasks. These results support our hypothesis that explicit alignment for clarification capabilities can significantly improve model performance on tasks requiring disambiguation, which is crucial for real-world programming scenarios where problem specifications are often ambiguous.}

\jw{ It should be noted that in this work, we focused on Comm. Rate and Good Question Rate in HumanEvalComm. Comm. Rate and Good Question Rate (in the first round) are indeed much more important metrics than the test pass rates and Pass@1 (in the second round). The reason is that: if an ideal LLM is given an ambiguous (or incomplete, or inconsistent) coding problem, and is instructed to ask questions or generate code, the LLM must ask questions ideally. The above results show that, to our knowledge, ClarifyCoder is one of the first models in the literature that is able to intelligently ask questions and know when not to ask questions. Table~\ref{tab:human_eval_results_compare} summarizes differences between our models and existing methods, such as Okanagan~\cite{wu2024benchmarking}, TICODER~\cite{lahiri2022interactive}, ClarifyGPT~\cite{10.1145/3660810}, ClariGen~\cite{miao2025clarigen}, which enforce user interaction for clarification. In contrast, ClarifyCoder exhibits a unique clarification-seeking behavior that more closely resembles human interaction.}



\begin{tcolorbox}[width=\linewidth,boxrule=0pt,top=1pt, bottom=1pt, left=1pt,right=1pt, colback=gray!20,colframe=gray!20]
\textbf{Answer to RQ1:} 
Our experiments demonstrate that clarify-aware fine-tuning substantially improves the original model' ability to ask effective clarifying questions, with ClarifyCoder achieving a communication rate of 63.61\% and a good question rate of 51.93\% (more than doubling the original model, i.e., DeepSeek Coder) while maintaining competitive code generation performance.

\end{tcolorbox}


\subsection{Breakdown on Clarification Categories}
\begin{table}[htbp]
\centering
\footnotesize
\begin{tabular}{@{}p{2.5cm}p{2.5cm}p{2.5cm}p{2.5cm}p{2.5cm}@{}}
\toprule
Model & Comm. Rate & Good Question Rate & \jw{2nd-Round Pass@1} & \jw{2nd-Round Test Pass Rate} \\
\midrule
\multicolumn{5}{c}{\textit{Category 1a (Ambiguous)}} \\
CodeLlama & 2.44\% & 2.13\% &14.02\% *** & 35.69\%\\
CodeQwen1.5  & 0.00\% & 0.00\% &46.34\%***&62.62\%*** \\
DS-Chat  & 23.78\% & 19.82\% &21.95\%**& 40.62\%**\\
DS-Coder  & 15.24\% & 13.41\% &48.05\%***&65.45\%*** \\
ClarifyCoder  & 62.80\% & 49.39\% &23.71\%***& 33.44\%***\\
\midrule
\multicolumn{5}{c}{\textit{Category 1c (Inconsistent)}} \\
CodeLlama  & 0.00\% & 0.00\% &30.67\% & 47.41\%\\
CodeQwen1.5  & 0.00\% & 0.00\% &67.68\% *& 79.90\%\\
DS-Chat  & 11.59\% & 8.23\% &39.63\%& 56.89\%\\
DS-Coder  & 1.83\% & 1.22\% &67.12\%**& 81.83\%**\\
ClarifyCoder  & 48.78\% & 36.59\% &35.9\% & 42.03\%\\
\midrule
\multicolumn{5}{c}{\textit{Category 1p (Incomplete)}} \\
CodeLlama  & 6.71\% & 6.40\% &9.76\% ***& 27.97\% ***\\
CodeQwen1.5  & 0.00\% & 0.00\% &46.95\%*** & 59.36\%*** \\
DS-Chat & 46.34\% & 38.72\% &21.95\%**& 43.73\%**\\
DS-Coder & 51.22\% & 46.95\% &32.33\%*** & 45.84\%*** \\
ClarifyCoder & 76.22\% & 67.07\% &32.41\%***& 48.51\%***\\
\midrule
\multicolumn{5}{c}{\textit{Category 2ac (Ambiguous + Inconsistent)}} \\
CodeLlama  & 1.85\% & 1.54\% &14.81\% ***& 34.87\% \\
CodeQwen1.5  & 0.00\% & 0.00\% &40.49\%*** & 59.80\%***\\
DS-Chat  & 22.22\% & 17.28\% &21.47\%**& 39.47\%***\\
DS-Coder  & 17.28\% & 14.51\% &39.44\%***& 55.79\%***\\
ClarifyCoder & 62.96\% & 50.00\% &20.83\%***& 31.46\%***\\
\midrule
\multicolumn{5}{c}{\textit{Category 2ap (Ambiguous + Incomplete)}} \\
CodeLlama  & 16.22\% & 14.19\% &8.45\%**& 22.35\%**\\
CodeQwen1.5  & 0.00\% & 0.00\% &28.57\%***& 44.96\%***\\
DS-Chat  & 45.95\% & 39.86\% &23.38\%& 35.89\%\\
DS-Coder  & 56.76\% & 50.68\% &36.84\%***& 49.09\%***\\
ClarifyCoder & 79.73\% & 69.59\% &34.92\%***& 47.26\%***\\
\midrule
\multicolumn{5}{c}{\textit{Category 2cp (Inconsistent + Incomplete)}} \\
CodeLlama  & 5.88\% & 4.41\% &8.57\%**& 33.41\%**\\
CodeQwen1.5  & 0.00\% & 0.00\% &37.14\%***&54.22\% ***\\
DS-Chat  & 35.29\% & 26.47\% &25.71\%& 48.22\%**\\
DS-Coder  & 5.88\% & 2.94\% &22.73\%***& 44.11\%***\\
ClarifyCoder & 55.88\% & 44.12\% &30.77\%***& 44.09\%***\\
\bottomrule
\multicolumn{5}{p{0.95\columnwidth}}{\footnotesize DS-Coder=DeepSeek Coder, DS-Chat=DeepSeek Chat. CodeQwen1.5=CodeQwen1.5 Chat. ClarifyCoder=ClarifyCoder w/ DeepSeek Coder as Base and loss on answers only. Test PR=Test Pass Rate, Comm.R=Communication Rate, Good Q.R=Good Question Rate. *p<0.1; **p\ensuremath{\leq}0.05; ***p<0.01}
\end{tabular}
\caption{Performance on HumanEvalComm by clarification category.}
\label{Tab:rq2}
\end{table}

\textbf{Breakdown on Categories with One Clarification Type.} Table~\ref{Tab:rq2} shows the results breakdown on the clarification categories 1a, 1c, and 1p, where only one level of clarification type (\textit{Ambiguity}, \textit{Inconsistency}, and \textit{Incompleteness}) is applied to the problem. For ClarifyCoder, among the three clarification types, \textit{Incompleteness} has the highest communication rate (76.22\%), surpassing the communication rates of \textit{Ambiguity} (62.80\%) and \textit{Inconsistency} (48.78\%). This indicates that \textit{Incompleteness} is relatively easier to detect and trigger clarification questions than \textit{Ambiguity} and \textit{Inconsistency}. A similar pattern is observed for DeepSeek Chat and DeepSeek Coder, where communication rates for \textit{Incompleteness} (46.34\% and 51.22\%) are significantly higher than for \textit{Inconsistency} (11.59\% and 1.83\%). \textit{Inconsistency} generally has the lowest communication rate among the three types, suggesting that it requires stronger reasoning capability to detect.
	
Good Question Rate follows similar patterns as the communication rate, with ClarifyCoder achieving the highest quality for \textit{Incompleteness} (67.07\%), followed by \textit{Ambiguity} (49.39\%) and \textit{Inconsistency} (36.59\%). This indicates that the quality of questions is proportional to the communication rate across different clarification types. Notably, ClarifyCoder performs much better than its base model DeepSeek Coder in \textit{Ambiguity} and \textit{Inconsistency} than in \textit{Incompleteness}, indicating the effectiveness of our method. 
	
However, CodeLlama and CodeQwen1.5 Chat exhibit different patterns. CodeLlama shows its highest communication rate for \textit{Incompleteness} (6.71\%), but CodeQwen1.5 Chat remains at 0\% communication rate across all three categories. This suggests  some Code LLMs, particularly CodeQwen1.5 Chat, are designed to prioritize code completion over clarification, even when requirements are incomplete or ambiguous. The high test pass rates indicate a potential data contamination issue, which seems more severe for CodeQwen1.5 Chat.


\jw{To compare different categories more comprehensively, we also report the pass rates in the 2nd round of HumanEvalComm evaluation. Besides the 1st round of HumanEvalComm evaluation, where Comm. Rates and Good Question Rates are calculated, the second round prompts the model to complete the code given the Q\&A results~\cite{wu2024benchmarking}. Thus, the second round of the HumanEvalComm evaluation is for code generation -- the input includes: 1) coding problem in the 1st round, 2) inquiry by ClarifyCoder (same for another model of interest), 3) answer to the inquiry (provided by LLM-Judge). If there is no inquiry in the 1st round, there will be no 2nd round and the generated code in the 1st round is recorded instead. See more details in~\cite{wu2024benchmarking}. } For the testing performance, \textit{Incompleteness} receives the lowest Pass@1 (9.76\% $\sim$ 46.95\%) and Test Pass Rate (27.97\% $\sim$ 59.36\%) across all models. This supports the hypothesis that without adequate clarification, incomplete problems lead to incorrect solutions. Conversely, \textit{Inconsistency} generally yields higher Pass@1 and Test Pass Rates, as models can sometimes generate correct code despite inconsistencies. For categories 1a, 1c, and 1p, most changes in Pass@1 and Test Pass Rate are statistically significant, with p-values below 0.01 for many comparisons.
	
\textbf{Breakdown on Categories with Two Clarification Types.} Table~\ref{Tab:rq2} also presents results for categories 2ac, 2ap, and 2cp, where combinations of two clarification types are applied. Compared to single clarification types, these combinations generally yield higher communication rates. For instance, ClarifyCoder's communication rate increases from 62.80\% for \textit{Ambiguity} alone to 79.73\% for \textit{Ambiguity} combined with \textit{Incompleteness} (2ap). Similarly, DeepSeek Coder's communication rate increases from 51.22\% for \textit{Incompleteness} alone to 56.76\% for \textit{Ambiguity} with \textit{Incompleteness} (2ap).

However, test performance metrics significantly decrease when moving from single to dual clarification types. 2cp (combining \textit{Inconsistency} and \textit{Incompleteness}) generally results in the lowest performance. The statistical significance of these differences is notable, with 75\% of changes in Pass@1 and Test Pass Rate having p-values below 0.05, and many below 0.01. This confirms that combining clarification types significantly increases problem difficulty, requiring more sophisticated clarification strategies.

\jw{It's important to note that: Pass@1 (same for Test Pass Rates) in HumanEval refers to code generation ability, while Pass@1 in HumanEvalComm means the Pass@1 in postclarification (code generation in 2nd round) after the LLM has engaged in clarification dialog and received answers to its clarifying questions. In HumanEvalComm evaluation, we directly use the ClarifyCoder as the model for both the 1st round and 2nd round. Thus, the ClarifyCoder is not fine-tuned or optimized for the 2nd round input, in which the model is given the Q\&A results as well as the coding task. In the second round, the same input format is used for all models to ensure fairness of evaluation. 
So it's expected that the pass rates may not be optimized in the 2nd round of the model - we just want to compare the difference between categories. After investigation, we found the reason is that: ClarifyCoder is fine-tuned specifically in the clarification round (1st round - task: clarification or code generation). This is causing it to get weaker results in the postclarification round (2nd round- task: pure code generation). Future work should include a more dedicated fine-tuning for 2nd round input as well to get the pass rates for HumanEvalComm. Nevertheless, we do observe higher pass rates of ClarifyCoder in \textit{Incomplete} category and \textit{"Inconsistent + Incomplete"} category. The reason could be that more useful information for incomplete problems than other categories is obtained in Q\&A results for generating correct code.} 

    
Overall, these results highlight that while ClarifyCoder maintains superior communication rates across all categories, the challenge of addressing multiple clarification types simultaneously remains substantial. The findings suggest that future work should focus on improving models' ability to handle complex combinations of clarification needs, particularly those involving inconsistencies.


\begin{tcolorbox}[width=\linewidth,boxrule=0pt,top=1pt, bottom=1pt, left=1pt,right=1pt, colback=gray!20,colframe=gray!20]
\textbf{Answer to RQ2:} 
\textit{Incompleteness} is easiest for models to detect while \textit{Inconsistency} is most challenging; combining clarification types increases communication rates but decreases code generation performance. ClarifyCoder demonstrates its effectiveness by consistently outperforming other models across all categories, in particular \textit{Ambiguity} and \textit{Inconsistency}.
\end{tcolorbox}

\subsection{Effectiveness of Synthetic Data Generation}

\begin{table}[h]
    \centering
    \caption{Perplexity and Entropy Analysis of Problems and Answers in Synthetic Data}
    \label{tab:perplexity_entropy}
    \begin{tabular}{lcc}
        \hline
        \textbf{Metric} & \textbf{Problems} & \textbf{Answers} \\
        \hline
        Average Perplexity & 26.10 & 17.64 \\
        Average Entropy & 4.06 & 3.47 \\
        \hline
    \end{tabular}
\end{table}

To answer RQ3, we analyzed perplexity and entropy of our synthetic data. Perplexity measures how well a language model can predict a text sequence—higher values indicate text that is more difficult to predict, suggesting greater ambiguity or complexity. Entropy quantifies the uncertainty in text prediction—higher values reflect greater variability and unpredictability in the content.

These metrics are particularly suitable for evaluating our synthetic data generation because they directly measure whether our modification process successfully introduced the intended ambiguity into problems while maintaining coherent answers.

Our analysis shows that modified problems have significantly higher perplexity (26.10) than answers (17.64), indicating that problems were successfully modified to be more ambiguous and difficult to predict. Similarly, the higher entropy in problems (4.06 vs. 3.47) confirms greater uncertainty and diversity in problem wording.

The lower metrics for answers suggest that, despite responding to ambiguous problems, clarifying questions maintained a more predictable and consistent structure—an important balance for effective training data.

\begin{tcolorbox}[width=\linewidth,boxrule=0pt,top=1pt, bottom=1pt, left=1pt,right=1pt, colback=gray!20,colframe=gray!20]
\textbf{Answer to RQ3:} 
Modified problems exhibited 54\% higher perplexity and 17\% greater entropy than answers, confirming effective ambiguity injection while preserving answer consistency.
\end{tcolorbox}

\subsection{Training Free Methods}
\begin{figure}[h] 
    \centering
    \includegraphics[width=0.6\columnwidth]{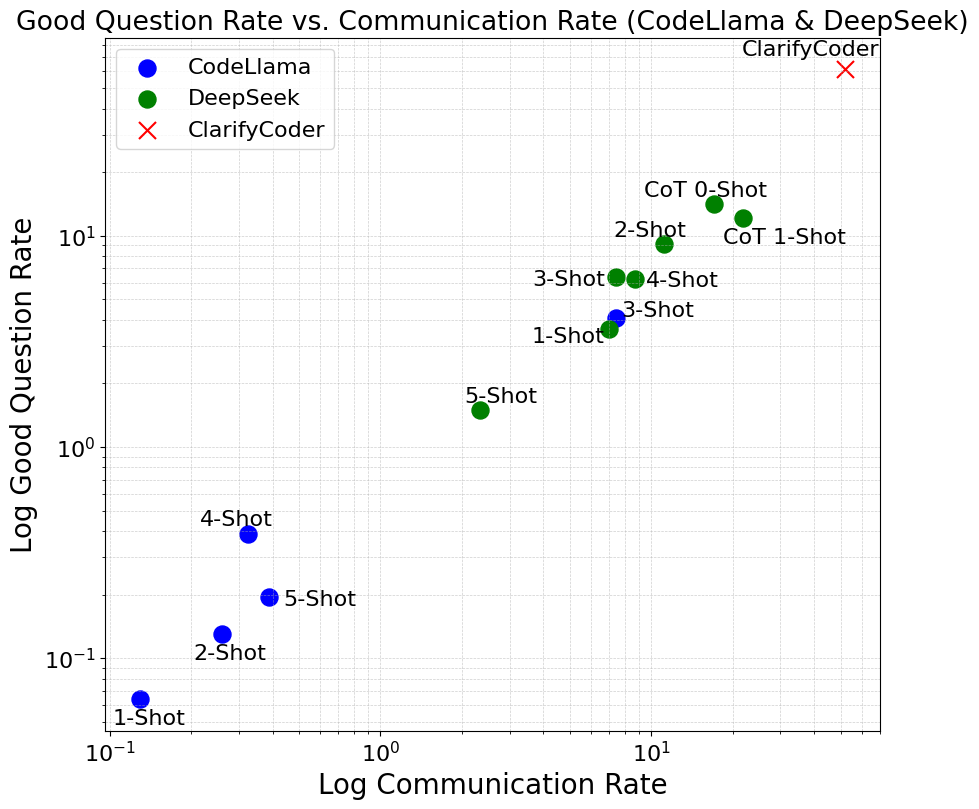} 
    \caption{Results on \textit{HumanEvalComm V2}  For In-Context Learning and Chain-of-Thought for CodeLLama and DeepSeek. The reported Communication rates and Good Question Rates are based on a log scale.}
    \label{fig:combined-in-context}
\end{figure}

Recently, there has been extensive studies conducted on in-context learning, which includes providing examples in the input prompt to the model, for it to try to make a more educated prediction. Compared to Supervised Fine-Tuning, or Parameter-Efficient Fine-Tuning, in-context learning does not require any additional parameter update of the model, thus making the overall procedure less resource-intensive. Another line of research has focused on Chain-of-Thought (CoT) prompting~\cite{wang2022towards}, where the model is asked to ``reason step-by-step''. This has proven to be an effective way to boost model performance in solving complex reasoning tasks. In order to stress the importance of clarify-aware fine-tuning, it is vital to compare its performance to that of in-context learning and CoT prompting to ensure that the increase communication capability of LLMs that we were able to achieve using clarify-aware fine-tuning would not be possible by prompting.
For that reason, we have conducted the following experiments: we have constructed prompts in a 1-shot, 2-shot, 3-shot, 4-shot, and 5-shot setting, meaning that within the prompt, there are examples of questions to which the model has to answer with either clarifying questions or a direct code snippet. We also do an evaluation using Chain-of-Thought prompting in a 0-shot and 1-shot settings. These experiments were conducted using CodeLlama and DeepSeek Coder models, and the corresponding results are summarized in Figure \ref{fig:combined-in-context}.

The results contain the LLM-based Communication Rate and the Good Question Rate on a log-scale on the HumanEvalComm V2 dataset.
Based on our evaluation, we can see that, for CodeLlama, adding more examples (shots) to the base prompt improves the communication rate and the good question rate at first. However, as we provide more and more examples, the overall performance starts to degrade, indicating that adding more examples is not always associated with getting better results. Interestingly enough, the CoT prompting resulted in 0 communication rate, which is why they do no appear on a log-scale plot. A similar pattern in few-shot prompting can be observed for DeepSeek Coder. Here, again, the communication rate starts to drop as we start adding 3 or more examples to the prompt. Nonetheless, CoT prompting produced more promising results in this case, as it surpassed all few-shot prompting baselines.
Even though these prompting techniques showed somewhat comparable performance to a simple instruction prompting, where we simply described the task to the model, they are still lagging behind the models that were clarify-aware fine-tuned. This shows the significance of our study as simple instruction prompting, In-context Learning, or Chain-of-Thought Reasoning were not able to provide the same increase in the communication competency as our approach in the base models.

\begin{tcolorbox}[width=\linewidth,boxrule=0pt,top=1pt, bottom=1pt, left=1pt,right=1pt, colback=gray!20,colframe=gray!20]
\textbf{Answer to RQ4: } Training-Free Methods, including few-shot prompting and CoT reasoning showed significantly lower communication competence compared to Clarify-Aware fine-tuning, signifying the effectiveness of our proposed approach.
\end{tcolorbox}

\begin{figure}[htbp]
\begin{center}
\begin{tikzpicture}
\begin{axis}[
    ybar,
    bar width=0.5cm,
    enlargelimits=0.15,
    enlarge y limits=false, 
    legend style={at={(0.5,-0.25)},
      anchor=north,legend columns=-1},
    ylabel={Rate},
    symbolic x coords={Comm. Rate, Good Q. Rate},
    xtick=data,
    nodes near coords,
    height=6cm, 
    ymin=0, 
    ymax=0.4, 
    ]
\addplot[fill=blue!30] coordinates {
    (Comm. Rate, 0.22) 
    (Good Q. Rate, 0.18)      
};
\addplot[fill=red!30] coordinates {
    (Comm. Rate, 0.23) 
    (Good Q. Rate, 0.20)      
};
\node[anchor=south, red] at (axis cs:Comm. Rate,0.3) {\footnotesize $\kappa=0.948$}; 
\node[anchor=north, red, yshift=20pt] at (axis cs:Good Q. Rate,0.20) {\footnotesize $\kappa=0.823$}; 
\legend{LLM-Based Metric, Manual Evaluation Metric}
\end{axis}
\end{tikzpicture}
\end{center}
\caption{Comparison of LLM-based and manual evaluation metrics for Comm. Rate and Good Q. Rate from 100 samples of ClarifyCoder responses. The inter-rater agreement for the manual evaluation is quantified using Kappa values ($\kappa$).}
\label{fig:manual}
\end{figure}

\section{Discussions}

\textbf{Manual Evaluation of LLM-based metrics.} To investigate the reliance of LLM-based metrics, we conducted a manual evaluation, by two authors using 100 samples from ClarifyCoder responses, to manually assess the clarify-awareness of models' responses and compare them with LLM-based metrics(Comm. Rate and Good Q. Rate). Figure~\ref{fig:manual} shows the comparison between human labeled metrics and LLM-based metrics. The Kappa is between 0.8-1.0, indicating that two raters reach substantial or near-perfect agreements. For both rates, LLM-based metrics align well with the human-labeled metrics, with less than 12\% differences. This shows the LLM-based metrics are accurate and aligns well with human judgment.

\textbf{Ratio of standard and synthetic data.}  In our investigation of the impact of synthetic data ratio on model performance, we experimented with different versions of ClarifyCoder, fine-tuning them with varying ratios of clarify-aware synthetic data to the entire data. The ratios ranged from 20\% to 100\%, implemented using downsampling. Our result in Figure~\ref{fig:ratio} that even with a modest 20\% ratio, utilizing just 2,191 clarify-aware samples, we observed a notable improvement in clarify-aware performance, with the Comm. Rate increasing by 6.45\% (from 36.90\% to 43.35\%) compared to the baseline. However, the best results were achieved when fine-tuning exclusively on clarify-aware synthetic data (100\% ratio). This leads to the best performance, with a substantial increase of 14.06\% in Comm. Rate (from 43.35\% to 57.41\%) and 10.77\% in Good Q. Rate (from 36.90\% to 47.67\%) compared to the combined fine-tuning with a 20\% ratio. These results show that combined fine-tuning improves clarify-awareness, but using exclusively the clarify-aware data gets the clarify-awareness performance.

\textbf{Calculating loss on answer only.} We also investigated the performance difference of ClarifyCoder on loss calculation during fine-tuning. We compared the clarify-aware fine-tuning where loss is calculated on both the problem and answer, and the clarify-aware fine-tuning where loss is calculated solely on the answer. As shown in Table \ref{tab:experiment_performance_results}, ClarifyCoder with loss calculated only on the answer achieves the highest communication rate (63.61\%) and good question rate (51.93\%), outperforming the version with loss calculated on both problem and answer. This improvement in clarification metrics comes with a slight trade-off in code generation performance, with Pass@1 and Test Pass Rate decreasing by 5.92 and 6.29 percentage points respectively. \FHF{However, Table \ref{tab:experiment_performance_results} reveals that }on the standard HumanEval benchmark, the answer-only loss version of ClarifyCoder actually performs better, with a Pass@1 of 74.85\% and Test Pass Rate of 83.29\%, compared to 73.01\% and 82.00\% for the version with loss on both problem and answer. 

\section{Threats of Validity}
\textbf{Internal Validity}.
This threat relates to the internal parameters such as the parameters in open-source Code LLMs. To mitigate this threat, we tested different parameters, such as fine-tuning on data of varying ratio, calculating the loss on both problems and answers, or solely on
the answers, optimizing the prompts as training-free methods and analyzed the results. 

\textbf{External Validity}. This threat relates to the effectiveness of the LLM-based metrics used in the evaluation. To mitigate this issue, we have conducted manual evaluation to assess the difference between LLM-based metrics and human labels. Besides, the LLM-based evaluator is used equally for all models in the evaluation, so this threat does not affect the relative ranking of the results for all models.

\textbf{Construct Validity}. This threat concerns the correctness of our generated synthetic data - While we have attempted to create realistic ambiguities and inconsistencies, the complexity of actual software requirements may not be fully represented in our synthetic dataset. We have analyzed the perplexity and entropy of synthetic data to mitigate this issue.

\begin{figure}[t]
\centering
\begin{tikzpicture}
\begin{axis}[
    width=0.6\textwidth,
    height=0.4\textwidth,
    xlabel={Ratio $r$ in ClarifyCoder},
    ylabel={Rate (\%)},
    xmin=20, xmax=100,
    ymin=20, ymax=70,
    xtick={20,40,60,100},
    grid=both,
    legend style={at={(0,1)}, anchor=north west, font=\tiny, draw=none, legend columns=2}, 
    thick,
    mark size=3pt,
    every axis plot/.append style={semithick},
    clip=false
]
\addplot[solid, color=black, mark=triangle*, line width=1.2pt] coordinates {
(20, 24.12) (100, 24.12)
};
\addplot[dashed, color=gray, mark=diamond*, line width=1.2pt] coordinates {
(20, 21.29) (100, 21.29)
};

\addplot[solid, color=red, mark=square*, line width=1.2pt] coordinates {
(20, 43.35) (40, 46.96) (60, 43.48) (100, 57.41)
};
\addplot[dashed, color=blue, mark=o, line width=1.2pt] coordinates {
(20, 36.90) (40, 40.52) (60, 38.12) (100, 47.67)
};

\legend{w/o ClarifyCoder (Comm. Rate), w/o ClarifyCoder (Good Q. Rate), w/ ClarifyCoder (Comm. Rate), w/ ClarifyCoder (Good Q. Rate)}
\end{axis}
\end{tikzpicture}
\caption{ClarifyCoder performance with varying ratio $r$.}
\label{fig:ratio}
\end{figure}

\section{Related Work}

\textbf{Clarify-Aware Code Generation with LLMs.} Large Language Models have demonstrated remarkable capabilities in generating code from natural language descriptions \cite{Li2022, Austin2021}. These models, trained on extensive code corpora, exhibit emergent capabilities in generating high-quality code solutions across various programming tasks \cite{Jiang2024a, Chen2023}. However, these LLMs often struggle with incomplete or underspecified instructions—a common scenario in software development \cite{Li2022, Chen2023}. This limitation is particularly pronounced in professional programming environments where requirements evolve and specifications may be initially vague \cite{Jiang2024a}. Several approaches address this challenge: ClarifyGPT \cite{10.1145/3660810} provides a framework that detects ambiguous requirements and generates targeted clarifying questions, demonstrating significant improvements in code generation performance across multiple benchmarks. ClariGen \cite{miao2025clarigen} integrates a clarifying Q\&A phase into the code generation process, enabling the LLM to produce more contextually informed and accurate code. TiCoder\cite{lahiri2022interactive} proposes a novel interactive workflow for guided intent clarification through tests. These approaches aim to bridge the gap between vague user requirements and precise code implementation. In this work, we argue that the ability to identify ambiguity and ask clarifying questions should be an intrinsic capability of the models themselves, so we study the alignment techniques such as fine-tuning with synthetic data to increase the model's clarify-awareness.

\textbf{Clarification and Alignment Approaches.} Asking clarifying questions (ACQ) is also studied and used in NLP tasks. ACQ is part of Question Generation (QG) for acquiring additional knowledge and resolving ambiguities from users' intent, as described in \cite{toles2023alexpaca,Shi2022, Zou2023}. Krasheninnikov et al. \cite{Krasheninnikov2022} fine-tuned language models on conversational data consisting of ambiguous user requests, clarifying questions, and final answers, demonstrating improved performance on addressing ambiguous instructions. Kuhn et al. \cite{Kuhn2022} showed that LLMs can reason about ambiguous aspects of a query and generate clarification questions with zero-shot prompting. Li et al. \cite{Li2023} proposed a framework in which LLMs infer intended behavior by querying the user with examples to label, yes-or-no questions, or open-ended questions. Their findings suggest that LLM-generated queries can be more efficient and require less effort than user-written prompts, enabling the discovery of initially unanticipated considerations of a task. Alexpaca introduces a novel task focused on generating factual clarification questions for multi-hop reasoning tasks without using examples \cite{toles2023alexpaca}.

\section{Conclusion}

In this research, we introduce ClarifyCoder, a novel approach to fine-tuning a Code LLM to learn the clarification-seeking behavior of recognizing ambiguity when it exists and requesting clarification before generating code. Through the combination of synthetic data generation and targeted instruction tuning, ClarifyCoder significantly improves clarification-awareness of a given LLM in code generation without sacrificing code generation abilities for well-specified coding tasks that don't need clarifications. By analyzing performance across different clarification categories and comparing with training-free baselines, we show that ClarifyCoder not only enhances clarify-awareness but also maintains efficiency and utility.  Furthermore, we show that optimizing both clarity-awareness and code generation capability could be conducted in a single fine-tuning with trade-offs. 

\bibliographystyle{ACM-Reference-Format}
\bibliography{references}

\end{document}

%% file: macros.tex
\newcommand{\anjiang}[1]{}